\documentstyle[12pt,prd,aps]{revtex}
\tightenlines
\draft
\begin{document}
\title{ Torsion, Scalar Field, Mass and FRW Cosmology}

\author{PRASANTA MAHATO\thanks{email : pmahato@dataone.in} }
\address{ Department of Mathematics, Narasinha Dutt College\\
         Howrah, West Bengal, India 711 101
         }
\date{\today}
\maketitle

\begin{abstract}
  In the Einstein-Cartan space $U_4$, an axial vector torsion together with a scalar field
  connected to a local  scale factor have been considered. By combining two
  particular
 terms from the $SO(4,1)$ Pontryagin  density and then modifying it in a $SO(3,1)$ invariant
 way, we get a Lagrangian density with Lagrange multipliers. Then under FRW-cosmological
  background, where the scalar field is connected to the source of gravitation, the
  Euler-Lagrange equations ultimately give the constancy of the gravitational constant
  together with only three kinds of energy densities representing mass, radiation and
  cosmological constant. The gravitational constant has been found to be linked with
  the geometrical Nieh-Yan density.
\end{abstract}
\maketitle
\section{Introduction}

At present standard cosmology starts with two basic assumptions:
(i) at sufficiently large scale matter distribution is spatially
homogeneous and isotropic and (ii) the large scale structure of
the universe can be described by Einstein's theory of gravity. The
geometrical evolution  of the universe can then be determined by
Einstein's equations where the energy momentum tensor acts as the
source. The Friedmann-Robertson-Walker(FRW)\cite{Fri22,Rob29,Wal35}
 universe is so far the most provocative and important
cosmological model of the universe. It is also one of the
simplest. It is isotropic, spatially homogeneous, and
fluid-filled. The FRW models serve as an introduction to the study
of homogeneous models. A FRW universe admits a six-parameter group
of isometries whose surfaces of transitivity are spacelike
three-surfaces of constant curvature. Minkowski space, de sitter
space and anti-de Sitter space are all special cases of the
general FRW spaces\cite{Haw73}. When several noninteracting
sources are present in the universe, the total energy momentum
tensor which appears on the right hand side of the Einstein's
equation will be the sum of the energy momentum tensor for each of
the sources.  Spatial homogeneity and isotropy imply that the
energy momentum tensor for the $i$-th source is diagonal and has
the form $ \mathop T\limits^{(i)}$$_{{}_{\beta}}{}^{\alpha}=$ dia
$[\rho_i,-p_i,-p_i,-p_i]$. Here $\rho_i$ and $p_i$ are
respectively the energy density and the pressure for the $i$-th
source which obey the energy conservation law $d(\rho_i
a^3)=-p_id(a^3)$, where $a(t)$ is the radius of the universe at
time $t$. The evolution of the energy density of each component is
essentially dependent on the parameter
$\omega_i\equiv\frac{p_i}{\rho_i}$. In particular $\omega_i$ $=$
$0$, $\frac{1}{3}$ or $-1$ respectively for non relativistic mass
density, radiation density or vacuum energy density\cite{Pad03}.

 It is well known that if we add the cosmological constant as the only source of
 curvature in Einstein's equation, the resulting space time is highly symmetric and
 has an interesting geometrical structure. In particular, in the case of positive
 cosmological constant, we get the well known de Sitter manifold\cite{Pad03}.

 Kibble\cite{Kib61} and Sciama\cite{Sci62} pointed out that the \textit{Poincar$\acute{e}$}
 group, which is the semi-direct product of translation and Lorentz rotation, is the
 underlying gauge group of gravity and found the so-called Einstein-Cartan theory
 where mass-energy of matter is related to the curvature and spin of matter is related
 to the torsion of space-time.
   One  major drawback of  \textit{Poincar$\acute{e}$}
  group is that it is a non-semisimple group which implies that there is no Lagrangian
  yielding its Yang-Mills equations\cite{Ald88}. There exists a general procedure\cite{Ald89}
 to check whether or not a set of field equations leads to a coherent theory, i.e. a theory
that can be quantized. If we apply it to Yang-Mills equations for non-semisimple groups,
we find that  they are never consistent. Here we see that though the
\textit{Poincar$\acute{e}$} group is the classical group for relativistic kinematics,
it cannot be given a quantum version. Now by minimal addition of extra terms this
inconsistent theory can be transformed to a good theory and  we   find a Lagrangian
of a gauge theory for a semi-simple group, the de Sitter group\cite{Ald88a}.
In this way, the de Sitter gauge theory comes up as the corrected
\textit{Poincar$\acute{e}$} gauge theory. Alternatively, there are other
approaches where de Sitter group based Yang-Mills theories are shown to be
producing either Ashtekar formulation of gravity\cite{Nie94} or
Einstein-Cartan version of general relativity\cite{Can02}.

    It is a remarkable result of differential geometry that
 certain global features of a manifold are determined by some local invariant densities.
These topological invariants have  an important property in common -  they are total divergences
and in any local theory  these invariants,  when treated as Lagrangian densities, contribute
 nothing to the Euler-Lagrange equations.  Hence in a local theory only few parts, not the
  whole part, of these invariants can be kept in a Lagrangian density. Recently, in
this direction, a gravitational Lagrangian has been
proposed\cite{Mah02a}, where a
 Lorentz invariant part of the de Sitter Pontryagin density has been treated as
 the Einstein-Hilbert Lagrangian.  By this way the role of torsion in the underlying manifold
has become multiplicative
   rather than additive one and  the  Lagrangian  looks like
    $torsion \otimes curvature$.  In other words - the additive torsion is decoupled from the
   theory but not the multiplicative one. This indicates that torsion is uniformly nonzero
   everywhere. In the geometrical sense, this implies that
   microlocal space-time is such that at every point there is a
   direction vector (vortex line) attached to it. This effectively
   corresponds to the noncommutative geometry having the manifold
   $M_{4}\times Z_{2}$, where the discrete space $Z_{2}$ is just
   not the two point space\cite{Con94} but appears as an attached direction vector.
In this paper we shall try
  to establish the `constancy' of this gravitational constant under the background of a
scalar field $\phi$
  which is either localized at laboratory scale or connected to the local  universal scale factor of an
   isotropic and homogeneous universe and, in particular,  also try to derive
    the power law of the cosmic energy density with respect to the local scale
    factor.

 \section{ Pontryagin density, Scalar field  and gravity Lagrangian}

 Cartan's structural equations for a Riemann-Cartan space-time $U_{4}$ are given by
\cite{Car22,Car24}
 \begin{eqnarray}T^{a}&=& de^{a}+\omega^{a}{}_{b}\wedge e^{b}\label{eqn:ab}\\
 R^{a}{}_{b}&=&d\omega^{a}{}_{b}+\omega^{a}{}_{c}\wedge \omega^{c}{}_{b},\label{eqn:ac}
\end{eqnarray}
here $\omega^{a}{}_{b}$ and e$^{a}$ represent the spin connection
and the local frames respectively.

In $U_{4}$ there exists  two invariant closed four forms. One is the well
known Pontryagin\cite{Che74,Che71} density \textit{P} and the
other is the less known Nieh-Yan\cite{Nie82} density \textit{N}
given by
\begin{eqnarray} \textit{P}&=& R^{ab}\wedge R_{ab}\label{eqn:ad}\\  \mbox{and} \hspace{2 mm}
 \textit{N}&=& d(e_{a}\wedge T^{a})\nonumber\\
&=&T^{a}\wedge T_{a}- R_{ab}\wedge e^{a}\wedge
e^{b}.\label{eqn:af}\end{eqnarray}

 The minimal Lagrangian density of a spin-$\frac{1}{2}$ field $\psi$, with an external
gravitational
 field with torsion, is given
by\cite{Mie01}\begin{eqnarray}L_{D}&=&\frac{i}{2}\{\overline{\psi}{}^* \gamma \wedge
D\psi+\overline{D\psi}\wedge{}^*\gamma\psi\}+{}^*m\overline{\psi}\psi-\frac{1}{4}A\wedge
\overline{\psi}\gamma_5{}^*\gamma\psi,\label{eqn:ayx}\end{eqnarray}where the
exterior covariant derivative $D$ is torsion-free,
$A$ is the axial vector part of the   torsion two form, $\gamma=\gamma_\mu dx^\mu=\gamma_ae^a$ and ${}^*$ is the Hodge duality
operator. Therefore,
considering the source in the matter Lagrangian, we can simply
assume that the torsion is given by an axial vector only.

In presence of
  axial vector  torsion, one naturally gets the Nieh-Yan
density from  (\ref{eqn:af})
\begin{eqnarray} N&=&-R_{ab}\wedge e^{a}\wedge e^{b}=-{}^* N\eta\hspace{2 mm},
\label{eqn:xaa}\\
 \mbox{where} \hspace{2 mm}\eta&:=&\frac{1}{4!}\epsilon_{abcd}e^{a}\wedge e^b\wedge
e^c\wedge e^d\end{eqnarray}is the invariant volume element.  It follows that    ${}^*N$,
the Hodge dual of $N$, is a scalar density of dimension $(length)^{-2}$.

We can combine the spin connection and the vierbeins multiplied by a scalar field together in a connection for $SO(5,1)$, in the tangent space, in the
form
\begin{eqnarray}W^{AB}&=&\left
[\begin{array}{cc}\omega^{ab}&\phi e^{a}\\- \phi e^{b}&0\end{array}\right],\label{eqn:aab}
\end{eqnarray}
where $a,b = 1,2,..4;A,B = 1,2,..5$ and $\phi$ is a variable parameter of dimension $(length)^{-1}$ and corresponds  a local length scale. In some earlier works\cite{Cha97,Mah02,Mah04}  $\phi$ has been treated as an inverse length constant.
With this connection we can obtain $SO(4,1)$ Pontryagin density as
\begin{eqnarray}F^{AB}\wedge F_{AB}&=&R^{ab}\wedge R_{ab}+2\phi^2d(e^a\wedge T_a)+4\phi d\phi\wedge e^a\wedge T_a\nonumber\\&=&P+dC_{T\phi},\label{eqn:xb}\end{eqnarray}where
\begin{eqnarray}C_{T\phi}&:=&2\phi^2 e^a\wedge T_a,\\
\textit{P}&:=&-R^{a}{}_b\wedge R^b{}_a=-(\bar{R}^{a}{}_b\wedge \bar{R}^b{}_a+2\bar{R}^{a}{}_b\wedge \hat{R}^b{}_a+\hat{R}^{a}{}_b\wedge \hat{R}^b{}_a),\label{eqn:xb1} \\ \bar{R}^b{}_a&=&d\bar{\omega}^b{}_a+\bar{\omega}^b{}_c\wedge \bar{\omega}^c{}_a,
\\\hat{R}^b{}_a&=&dT^b{}_a+\bar{\omega}^b{}_c\wedge T^c{}_a+T^b{}_c\wedge \bar{\omega}^c{}_a+T^b{}_c\wedge T^c{}_a \\\mbox{and}\hspace{5mm}T^{a}{}_{b}&=&\omega^{a}{}_{b}-\bar{\omega}^{a}{}_{b} \hspace{4mm}\mbox{s. t.}\hspace{5mm}T^{a}{}_{b}\wedge e^b=T^a\label{eqn:xb4}
\end{eqnarray}
Now $-\bar{R}^{a}{}_b\wedge \bar{R}^b{}_a$, the purely Riemannian torsion-less part of $P$,   is a closed four form and is given by
\begin{eqnarray}
-\bar{R}^{a}{}_b\wedge \bar{R}^b{}_a&=& -d(\bar{\omega}^a{}_b\wedge \bar{R}^b{}_a-\frac{1}
{3}\bar{\omega}^a{}_b\wedge\bar{\omega}^b{}_c\wedge\bar{\omega}^c{}_a)=dC_R\label{eqn:xb2}\\\mbox{where}\hspace{2mm}C_R&=&-(\bar{\omega}^a{}_b\wedge \bar{R}^b{}_a-\frac{1}
{3}\bar{\omega}^a{}_b\wedge\bar{\omega}^b{}_c\wedge\bar{\omega}^c{}_a)\nonumber.
\end{eqnarray}
With the hypothesis that only the axial vector part of the torsion is present in the physical world, we can write
\begin{eqnarray}
    T^a&=&e^{a\mu}T_{\mu\nu\alpha}dx^\nu\wedge dx^\alpha, \hspace{2mm}T^{ab}=e^{a\mu}e^{b\nu}T_{\mu\nu\alpha}  dx^\alpha\nonumber\\\hspace{2mm}\mbox{and}\hspace{2mm}{}^*A&=&T=\frac{1}{3!}T_{\mu\nu\alpha}dx^\mu\wedge dx^\nu\wedge dx^\alpha\hspace{4mm}\mbox{s.t.}\hspace{4mm}N=6dT
\end{eqnarray}
In this framework we see that
\begin{eqnarray}
    \hat{R}^a{}_b\wedge\hat{R}^b{}_a&=&-2d(A\wedge dA-\frac{1}{3}T^a{}_b\wedge T^b{}_c\wedge T^c{}_a)=-dC_T\\\mbox{and}\hspace{4mm}2\bar{R}^a{}_b\wedge\hat{R}^b{}_a&=&-4\mbox{$\mathcal{ R}$}dT+8\mbox{$\mathcal{ R}$}^{ab}\bar{\nabla}( A_b\eta_a)=8d(G^{ab}A_b\eta_a)=-dC_{RT}\\\mbox{where}\hspace{4mm}\eta_a&=&\frac{1}{3!}\epsilon_{abcd}e^b\wedge e^c\wedge e^d,\hspace{2mm}C_T=2(A\wedge dA-\frac{1}{3}T^a{}_b\wedge T^b{}_c\wedge T^c{}_a)\nonumber\\\mbox{and}\hspace{2mm}C_{RT}&=&-8(G^{ab}A_b\eta_a).\nonumber
\end{eqnarray}Here $\bar{\nabla}$ is the torsion-free covariant derivative; $\mathcal{R}$,  $\mathcal{R}$${}^{ab}$ and $G^{ab}$ are, respectively, corresponding Ricci scalar,  Ricci tensor and Einstein's tensor.

Hence we see that the $SO(4,1)$ Pontryagin density in $U_4$ is the sum of four closed four forms, given by
\begin{eqnarray}
F^{AB}\wedge F_{AB}&=&dC_R+dC_T+dC_{RT}+dC_{T\phi}.
\end{eqnarray}Since all these four forms are total divergences, they yield nothing in any local theory when treated as Lagrangian densities. Hence to have
an effective field theory, however,  we may consider   some Lorentz invariant parts of them as Lagrangian densities.  So here we heuristically propose a Lagrangian density which combines a part of $dC_{RT}$  with a part of $dC_{T\phi}$ as follows
\begin{eqnarray}\mbox{$\mathcal{L}$}_{0}= (\mbox{$\mathcal{R}$}-\beta\phi^2)dT=-\frac{1}{6}(\mbox{$\mathcal{R}$}-\beta\phi^2){}^*N\eta\end{eqnarray}
where $\beta$ is a dimensionless coupling constant.

   So far $SO(3,1)$
invariance is concerned, torsion can be separated from the
connection  as the torsional part of the $SO(3,1)$ connection
transforms like a tensor i.e. when vierbeins also transform like
$SO(3,1)$ tensors in a broken $SO(4,1)$ gauge theory. In this
direction it is important to define a torsion-free covariant
differentiation through a field equation involving the connection
and the vierbeins only. So  we introduce   Lagrangian
density $\mathcal{L}\mbox{${}_{1}$}$, given by,
\begin{eqnarray}\mathcal{L}\mbox{$_{1}$}=\mbox{${}^*(b_a\wedge
\bar{\nabla} e^{a})(b_a\wedge
\bar{\nabla} e^{a})$} ,\end{eqnarray} where   $\bar{\nabla}$
represents covariant differentiation with respect to a $SO(3,1)$
connection one form $\bar{\omega}^{ab}$ and $b_{  a}$ is a two form with
one internal index and of dimension $(length)^{-1}$. If we treat $b_{a}$ as Lagrange
multiplier
then it ensures that $\bar{\nabla}$ represents torsion-free covariant
differentiation. By this way torsion has become decoupled from the connection part of the theory. It has become independent of the one form $e^a$, in particular, owing to its fundamental existence as a metric independent tensor in the affine connection in $U_4$, we treat here the three form $T=\frac{1}{3!}e^a\wedge T_a$ as more fundamental than the one form $T^{ab}=\omega^{ab}-\bar{\omega}^{ab}$.\footnote{One may raise the aesthetic question of identifying $T$ with the torsion. This can be properly addressed if we introduce two separate $SO(3,1)$ connections $\omega^{ab}$ and $\bar{\omega}^{ab}$ and replace the Lagrangian $\mathcal{L}\mbox{$_{2}$}$ by the gauge invariant expression ${}^*[b_a\wedge(
\mathop \nabla\limits^\omega e^{a}-T^a)][b_a\wedge(
\mathop \nabla\limits^\omega e^{a}-T^a)]$+${}^*[c_a\wedge(
\omega^{ab}-\bar{\omega}^{ab}-T^{ab})][c_a\wedge(
\omega^{ab}-\bar{\omega}^{ab}-T^{ab})]$ where the three form $c_a$ is another Lagrange multiplier of proper dimension and $\mathop \nabla\limits^\omega$ is covariant differentiation w.r.t. the connection $\omega$.}

Now we add another Lagrangian density $\mathcal{L}$${}_{2}$ containing   a nonlinear
kinetic term,  given by
\begin{eqnarray}
    \mathcal{L}\mbox{${}_{2}=- f(\phi)d\phi\wedge{}^*d\phi -h(\phi)\eta$}
\end{eqnarray}where $f(\phi)$ and  $h(\phi)$ are unknown functions of $\phi$ whose
forms are to be determined subject to the geometric structure of the manifold.

  At last we are in a position to define the
total gravitational Lagrangian density in empty space, as,
\begin{eqnarray}\mbox{$\mathcal{L}$}_{G}&=&\mbox{$\mathcal{L}$}_{0} +
\mbox{$\mathcal{L}$}_{1} +\mbox{$\mathcal{L}$}_{2}
 ,\nonumber\\
&=& -\frac{1}{6}({}^*N \mbox{$\mathcal{R}$}\eta  +\beta \phi^2 N)+{}^*(b_a\wedge
\bar{\nabla} e^{a})(b_a\wedge
\bar{\nabla} e^{a}) ,\nonumber\\&{}&- f(\phi)d\phi\wedge{}^*d\phi -h(\phi)
\eta,\label{eqn:abcd1}\end{eqnarray} where
  *   is Hodge duality operator,   $N=6dT$, $\mathcal {R}$$\eta=\frac{1}{2}\bar{R}^{ab}\wedge\eta_{ab}$ and $\eta_{ab}={}^*(e_a\wedge e_b)$.    To
start with this Lagrangian we have altogether 69 independent
components of the field variables $e^{a}$, $T$,
 $\bar{\omega}^{ab}$, $\phi$ and   $b^{a}$.
  The geometrical implication of the first term, i.e. the $torsion \otimes curvature$\footnote{An important advantage of this part of the Lagrangian is that - it is a
   quadratic one with respect to the field derivatives and this
   could be valuable in relation to the quantization program of gravity like other gauge theories of
   QFT.} term, in the Lagrangian $\mathcal{L}\mbox{$_{G}$}$   has been already discussed in section one.

  \section{   Euler-Lagrange equations and gravitational constant  }

  The Lagrangian $\mathcal{L}\mbox{$_{G}$}$, which is defined in the previous section,
     is  only Lorentz invariant  under rotation in the tangent space where  de Sitter
     boosts are not permitted. As a consequence $T$ can be treated independently of $e^a$
     and $\bar{\omega}^{ab}$. Then following reference \cite{Heh95}, we independently vary
     $e^{a}$, $\bar{\nabla} e^{a}$, $dT$,  $\bar{R}^{ab}$, $\phi$, $d\phi$ and
    $b^{a}$  and find

\begin{eqnarray}
    \delta  \mathcal{L}\mbox{$_{G}$}&=&\delta e^a\wedge \frac{\partial
    \mathcal{L}\mbox{$_{G}$}}{\partial e^a}+\delta \bar{\nabla} e^a\wedge \frac{\partial
    \mathcal{L}\mbox{$_{G}$}}{\partial \bar{\nabla} e^a}+\delta dT \frac{\partial  \mathcal{L}\mbox{$_{G}$}}{\partial
    dT}+\delta \bar{R}^{ab}\wedge \frac{\partial  \mathcal{L}
\mbox{$_{G}$}}{\partial
    \bar{R}^{ab}}\nonumber\\&{}&+\delta\phi\frac{\partial  \mathcal{L}\mbox{$_{G}$}}{\partial \phi}+\delta
    d\phi\wedge\frac{\partial  \mathcal{L}\mbox{$_{G}$}}{\partial d\phi}+\delta b^a\wedge
    \frac{\partial  \mathcal{L}\mbox{$_{G}$}}{\partial b^a} \\&=&\delta e^a\wedge( \frac{\partial
    \mathcal{L}\mbox{$_{G}$}}{\partial e^a}+ \bar{\nabla}\frac{\partial  \mathcal{L}
    \mbox{$_{G}$}}{\partial \bar{\nabla} e^a})+\delta T\wedge d \frac{\partial  \mathcal{L}
    \mbox{$_{G}$}}{\partial dT}+\delta \bar{\omega}^{ab}\wedge(\bar{\nabla} \frac{\partial
     \mathcal{L}\mbox{$_{G}$}}{\partial \bar{R}^{ab}}+ \frac{\partial
    \mathcal{L}\mbox{$_{G}$}}{\partial \bar{\nabla} e^a}\wedge e_b)\nonumber\\&{}&+\delta \phi(\frac{\partial
    \mathcal{L}\mbox{$_{G}$}}{\partial \phi}- d \frac{\partial  \mathcal{L}\mbox{$_{G}$}}
    {\partial d\phi})+\delta b^a\wedge \frac{\partial  \mathcal{L}\mbox{$_{G}$}}{\partial
    b^a}\nonumber \\&{}&
    +d(\delta e^a\wedge \frac{\partial  \mathcal{L}\mbox{$_{G}$}}{\partial
    \bar{\nabla} e^a}+\delta T \frac{\partial  \mathcal{L}\mbox{$_{G}$}}{\partial dT}+\delta
    \bar{\omega}^{ab}\wedge \frac{\partial  \mathcal{L}\mbox{$_{G}$}}{\partial \bar{R}^{ab}}+
    \delta\phi\frac{\partial  \mathcal{L}\mbox{$_{G}$}}{\partial d\phi})\label{eqn:abc0}
\end{eqnarray}Using the form of the Lagrangian $\mathcal{L}\mbox{$_{G}$}$, given in (\ref{eqn:abcd1}),
we get
\begin{eqnarray}
    \frac{\partial  \mathcal{L}\mbox{$_{G}$}}{\partial e^a}&=&-\frac{1}{6}{}^*N(2\textbf{R}_a-\mbox{$\mathcal{R}$}\eta_a)
  -{}^*(b_b\wedge
\bar{\nabla} e^{b})^2\eta_a\nonumber\\&{}&-f(\phi)[-2\partial_a\phi\partial^b\phi\eta_b+\partial_b\phi\partial^b\phi\eta_a]-h(\phi)\eta_a\label{eqn:abc1} \\
\frac{\partial  \mathcal{L} \mbox{$_{G}$}}{\partial (\bar{\nabla}
e^a)}&=&2{}^*(b_a\wedge \bar{\nabla} e^{a})b_a\label{eqn:abc2}
 \\\frac{\partial  \mathcal{L}\mbox{$_{G}$}}{\partial (dT)}&=&\mbox{$\mathcal{R}$}-\beta\phi^2\label{eqn:abc4}\\\frac{\partial  \mathcal{L}\mbox{$_{G}$}}{\partial \bar{R}^{ab}}&=&-\frac{1}{24}{}^*N\epsilon_{abcd}e^c\wedge e^d=-\frac{1}{12}{}^*N\eta_{ab}\label{eqn:abc5}\\\frac{\partial  \mathcal{L}\mbox{$_{G}$}}{\partial \phi}&=&-\frac{1}{3}\beta \phi N- f^\prime(\phi)d\phi\wedge{}^*d\phi -h^\prime(\phi)
\eta\label{eqn:abc5a}\\\frac{\partial  \mathcal{L}\mbox{$_{G}$}}{\partial d\phi}&=&-2f{}^*d\phi\label{eqn:abc5b}\\\frac{\partial  \mathcal{L}\mbox{$_{G}$}}{\partial b^a}&=&2{}^*(b_b\wedge
\bar{\nabla} e^{b})
\bar{\nabla} e_{a}\label{eqn:abc6} \\
\mbox{Where} \hspace{10mm}&{}&\\
    \textbf{R}_a&:=&\frac{1}{2}\frac{\partial (\mbox{$\mathcal{R}$}\eta)}{\partial e^a}=\frac{1}{4}\epsilon_{abcd}\bar{R}^{bc}\wedge  e^d
\end{eqnarray}and ${}^\prime$ represents derivative w.r.t. $\phi$.

From above, Euler-Lagrange equations for  $b_a$ gives us
\begin{eqnarray}
\bar{\nabla} e_{a}&=&0\label{eqn:abc9}
\end{eqnarray}i.e.   $\bar{\nabla}$ is torsion free. Using this result in (\ref{eqn:abc1}) and  (\ref{eqn:abc2})  we get
\begin{eqnarray}
    \frac{\partial  \mathcal{L}\mbox{$_{G}$}}{\partial e^a}&=&-\frac{1}{6}{}^*N(2\textbf{R}_a-\mbox{$\mathcal{R}$}\eta_a)-f(\phi)[-2\partial_a\phi\partial^b\phi\eta_b+\partial_b\phi\partial^b\phi\eta_a]-h(\phi)\eta_a\label{eqn:abc10}\\    \frac{\partial  \mathcal{L}\mbox{$_{G}$}}{\partial (\bar{\nabla} e^a)}&=&0\label{eqn:abc11}
\end{eqnarray}Hence Euler-Lagrange equations of $e^a$, $T$ and $\bar{\omega}^{ab}$,using (\ref{eqn:abc0}),  (\ref{eqn:abc4})  and (\ref{eqn:abc5}) give us
\begin{eqnarray}\frac{1}{6}    {}^*N(2\textbf{R}_a-\mbox{$\mathcal{R}$}\eta_a)+f(\phi)[-2\partial_a\phi\partial^b\phi\eta_b+\partial_b\phi\partial^b\phi\eta_a]+h(\phi)\eta_a&=&0\label{eqn:abc13}\\d(\mbox{$\mathcal{R}$}-\beta\phi^2)&=&0\label{eqn:abc14}\\\bar{\nabla}({}^*N\eta_{ab})&=&0\label{eqn:abc15}
\end{eqnarray}
From (\ref{eqn:abc5a}) and (\ref{eqn:abc5b}), the Euler-Lagrange equations for the field $\phi$ is given by
\begin{eqnarray}
      -\frac{1}{3}\beta \phi N+ f^\prime(\phi)d\phi\wedge{}^*d\phi -h^\prime(\phi)
\eta+2fd{}^*d\phi&=&0.\label{eqn:abc15a}
\end{eqnarray}
 Using (\ref{eqn:abc9}) in (\ref{eqn:abc15})
\begin{eqnarray}
    d{}^*N=0\label{eqn:abc16}
\end{eqnarray}
From  equations (\ref{eqn:abc14}) and (\ref{eqn:abc16}) we can write
\begin{eqnarray}
{}^*N=\frac{6}{\kappa}\hspace{4mm}\mbox{and}\hspace{4mm} \mbox{$\mathcal{R}$}-\beta\phi^2=\lambda \label{eqn:abc17}
\end{eqnarray}where $\kappa$ and $\lambda$ are integration constants having dimensions of $(length)^{2}$ and $(length)^{-2}$ respectively.
Then using  properties $e^a\wedge\eta_b=\delta^a_b\eta$ and $ \textbf{R}_a=-G^b{}_a\eta_b$ where $G^b{}_a:=\mbox{$\mathcal{R}$}^b{}_a-\frac{1}{2}\mbox{$\mathcal{R}$}\delta^b{}_a$  in  (\ref{eqn:abc13}), we get
\begin{eqnarray}
    \textbf{R}_a=\kappa[f\partial_a\phi\partial^b\phi+\frac{h}{2}\delta^b{}_a]\eta_b, \label{eqn:abc18}
\end{eqnarray}such that,
\begin{eqnarray}
G^b{}_a&=&-\kappa[f\partial_a\phi\partial^b\phi+\frac{h}{2}\delta^b{}_a],\label{eqn:abc18a}\\\mbox{and}\hspace{4mm} \mbox{$\mathcal{R}$}\eta&=&\kappa[fd\phi\wedge{}^*d\phi+2h\eta].\label{eqn:abc19}\end{eqnarray}
 From (\ref{eqn:abc17}) and (\ref{eqn:abc19}) we get
\begin{eqnarray}[\frac{1}{\kappa}(\beta\phi^2+\lambda)-2h]\eta&=& fd\phi\wedge{}^*d\phi\nonumber\\ &=&(f\partial_c\phi\partial^c\phi)\eta\label{eqn:abc19a}
\end{eqnarray}
Eliminating $d\phi\wedge{}^*d\phi$ from  (\ref{eqn:abc15a}) and (\ref{eqn:abc19a})we get
\begin{eqnarray}
  \frac{2}{\kappa}\beta \phi \eta+ \frac{f^\prime}{f}[\frac{1}{\kappa}(\beta\phi^2+\lambda)-2h]\eta -h^\prime(\phi)
\eta+2fd{}^*d\phi&=&0.\label{eqn:abc21}
\end{eqnarray}
\section{  $\phi$ is localized at laboratory  scale }
Here we study the case where $\phi$ is a local scalar field which vanishes at space infinity
  and  has a quadratic Lagrangian. So we assume $f=\frac{1}{2}$, $\beta=\frac{c^2_\phi}{2}$
  and $h=constant$ in (\ref{eqn:abcd1}), where $c^2_\phi$ is the
  dimensionless $torsion\times\phi$ coupling constant, and then  (\ref{eqn:abc18a}) and
  (\ref{eqn:abc21}) reduce to,
\begin{eqnarray}
G^b{}_a&=&\kappa [\frac{1}{2} \partial_a\phi\partial^b\phi+\frac{h}{2}\delta^b{}_a],\label{eqn:abc22}\\
 d{}^*d\phi&=&-\frac{1}{\kappa}c^2_\phi\phi\eta.\label{eqn:abc23}
\end{eqnarray}Using the boundary condition of $\phi$ at space infinity on(\ref{eqn:abc19a}) we get
\begin{eqnarray}
     d\phi\wedge{}^*d\phi&=&\frac{1}{\kappa}c^2_\phi\phi^2 \eta.\label{eqn:abc24}
\end{eqnarray}where $\lambda=2h\kappa$. Equation (\ref{eqn:abc23}) is the  correct  field equation of a massive scalar field $\phi$ of mass $m_\phi$, provided, we define the mass by the following equation
\begin{eqnarray}
    m_\phi&=&\frac{c_\phi}{\sqrt{\kappa}}.\label{eqn:abc25}
\end{eqnarray}
This last equation shows that, through the NY-term, torsion is not only connected to the gravitational constant, it also gives mass of a scalar field through the $torsion\times\phi$ interaction term. Hence   the  gravitational constant and the mass of a scalar field have the same geometrical origin in the Riemann-Cartan space $U_4$.

\section{  $\phi$ and FRW  cosmology }
Here we study the case where $\phi$ represents the local energy scale in the back ground of FRW cosmology. In this back ground we assume $\phi$ to be a variable function of time only. Then, w. r. t. external indices, (\ref{eqn:abc18a}) becomes
\begin{eqnarray}
    G_{00}=-\kappa ( f\dot{\phi}^2  +\frac{h}{2}g_{00})\nonumber\\
    G_{ij}=-\kappa ( \frac{h}{2}g_{ij})\hspace{2mm}\mbox{where}\hspace{2mm}i,j=1,2,3.
\label{eqn:xp1}
\end{eqnarray}
Here   we shall try to solve   equations (\ref{eqn:abc19a}) and (\ref{eqn:abc21}) under the
isotropic and homogeneous  cosmological background of a   universe where the metric tensor
is given by the FRW metric
\begin{eqnarray}
 g_{00}=-1 ,\hspace{2mm}g_{ij}= \delta_{ij}a^2(t)  \hspace{2mm}\mbox{where}\hspace{2mm}
i,j=1,2,3;\label{eqn:aay}
\end{eqnarray}such that
\begin{eqnarray}
    &{}&e=\sqrt{-\det(g_{\mu\nu})}= a^3\label{eqn:zaz}
\end{eqnarray}With this assumption equation (\ref{eqn:abc19a}) reduces to
\begin{eqnarray}
     f \dot{\phi}^2=-\frac{1}{\kappa}(\beta\phi^2 +\lambda)+2h .\label{eqn:aaj}
\end{eqnarray}
   Now, with the cosmological restriction on the metric  as stated in  (\ref{eqn:aay})
 and the $\phi$-field is a function of time only, the equation (\ref{eqn:abc15a}) reduces to
\begin{eqnarray}
     2f \ddot{\phi}+2f\frac{e^\prime}{e}\dot{\phi}^2+ f^\prime\dot{\phi}^2-\frac{2\beta}
{\kappa}\phi + h^\prime=0
\end{eqnarray}If we eliminate $\ddot{\phi}$ from this equation with the help of the time derivative of equation
(\ref{eqn:aaj}),  we get
\begin{eqnarray}
 2f\frac{e^\prime}{e}\dot{\phi}^2&=&\frac{4\beta}{\kappa}\phi-3h^\prime \nonumber\\
 \mbox{or,}\hspace{5mm} 2\frac{e^\prime}{e}&=&-\frac{\frac{4\beta}{\kappa}\phi-3h^\prime}{\frac{1}{\kappa}(\beta\phi^2 +\lambda)-2h}\label{eqn:aal}
\end{eqnarray}
 Now, for the FRW metric, the non-vanishing components of Einstein's tensor (\ref{eqn:xp1})
are given by
\begin{eqnarray}
    G^{0}{}_{0}&=&-3(\frac{\dot{a}}{a})^2=-\kappa(\frac{\beta}{\kappa}\phi^{2}+\frac{\lambda}
{\kappa}-\frac{3h}{2}) \nonumber\\G^{j}{}_{i}&=&-(\frac{2\ddot{a}}{a}+\frac{\dot{a}^2}{a^2})
\delta^{j}{}_{i}=-\kappa\frac{h}{2}\delta^{j}{}_{i}\label{eqn:11a}
\end{eqnarray}Positive energy condition implies both $\beta$ and $\lambda$ are positive constants and from the forms of $G^{0}{}_{0}$ and $G^{j}{}_{i}$ it appears that the term $\frac{\beta}
{\kappa}\phi^{2}$ represents pressure-less energy density i.e. $\phi^{2}\propto a^{-3}\propto
 \frac{1}{e}$. Putting this in (\ref{eqn:aal}) we get after integration
\begin{eqnarray}
    h=-\gamma \phi^{\frac{8}{3}}+\frac{\lambda}{2\kappa}
\end{eqnarray} where $\gamma$ is a constant of dimension $(length)^{-\frac{4}{3}}$ . Using this
 functional form  of $h$ in (\ref{eqn:aaj}) and (\ref{eqn:11a}) we get
\begin{eqnarray}
f=-\frac{2}{3\kappa}\frac{F^\prime}{\phi F}\hspace{2mm}\mbox{where}\hspace{2mm}  F(\phi)=
\beta\phi^2+\frac{3}{2}\gamma\kappa\phi^{\frac{8}{3}}+\frac{\lambda}{4}
\end{eqnarray}
\begin{eqnarray}
G^{0}{}_{0}&=&-3(\frac{\dot{a}}{a})^2=-\kappa(\frac{\beta}{\kappa}\phi^{2}+\frac{3\gamma}
{2}\phi^{\frac{8}{3}}+\frac{\lambda}{4\kappa}) \nonumber\\G^{j}{}_{i}&=&-(\frac{2\ddot{a}}
{a}+\frac{\dot{a}^2}{a^2})\delta^{j}{}_{i}=\kappa(\frac{\gamma}{2}\phi^{\frac{8}{3}}-
\frac{\lambda}{4\kappa})\delta^{j}{}_{i}\label{eqn:22a}
\end{eqnarray} This form of $G^{0}{}_{0}$ and $G^{j}{}_{i}$ implies that, in the present
framework, at cosmic scale, only three types of energy densities are possible, viz.
\begin{enumerate}
\item The pressure-less mass density $\rho_{{}_M}=\frac{\beta}{\kappa}\phi^{2}\propto
a^{-3}$,
\item The radiation density $\rho_{{}_R}=\frac{3\gamma}{2}\phi^{\frac{8}{3}}\propto
a^{-4}$ where pressure $p_{{}_R}=\frac{1}{3}\rho_{{}_R}$ and
\item The constant vacuum energy density $\rho_{{}_{VAC.}}=\frac{\lambda}{4\kappa}$
where pressure $ p_{{}_{VAC.}}=-\rho_{{}_{VAC.}}$,
\end{enumerate}where $\beta$, $\gamma$ and $\lambda$ are all positive constants.
 Hence we can write
\begin{eqnarray}
        G_{00}&=&3H^2= \kappa\rho ,\nonumber\\
        G_{ij}&=& \kappa pa^2\delta_{ij} \hspace{2mm}\mbox{where}\hspace{2mm}i,j=1,2,3;
\label{eqn:zz2}
\end{eqnarray}where the Hubble's parameter $H=\dot{a}/a$, $\rho=\rho_{{}_M}+\rho_{{}_R}+
\rho_{{}_{VAC.}}$ and $p=p_{{}_R}+p_{{}_{VAC.}}$, such that $\rho$ obeys, as  a consequence
of Bianchi identity $G^\mu{}_0{}_{;\mu}=0$, the energy conservation law of Newtonian
mechanics,     given by the equation of state\cite{Car01,Pad03}
\begin{eqnarray}
    d( \rho a^3)=-p d(a^3). \label{eqn:zz3}
\end{eqnarray}
Now from (\ref{eqn:22a}) and  (\ref{eqn:zz2}), we get after
eliminating $(\frac{\dot{a}}{a})^2$ that
\begin{eqnarray}
     \frac{\ddot{a}}{a}=-\frac{\kappa}{6}(\rho+3p)\label{eqn:zz4}.
\end{eqnarray}Equations (\ref{eqn:zz2}) and (\ref{eqn:zz4}) are two well known results
of FRW cosmology\cite{Wal84}. Hence in this  background, where  de Sitter gauge symmetry
is broken in a Lorentz invariant way linking gravitational constant with the NY density, we have found FRW cosmology  with only three kinds
of energy density.  Two of
these kinds are that of a perfect fluid where $p=0$,  $\frac{\rho}{3}$ and the remaining
type is that of vacuum energy where $p=-\rho$. At first glance this result looks nothing new. In standard model $\rho=\rho_{{}_M}+\rho_{{}_R}+\rho_{{}_{VAC.}}$ is assumed empirically but other forms of energy densities are not ruled out subject to the pressure-energy relation (\ref{eqn:zz3}). But in our present formalism other forms of energy densities imply different forms of the functions $f$ and $h$ as solutions and this indirectly implies departure from the FRW metric at the cosmic scale. This is not the case we are studying here.

Hence the differential equation of the
evolution of the universe can be written from (\ref{eqn:zz4}) as
\begin{eqnarray}
    \frac{\ddot{a}}{a}
    &=& -\frac{\kappa}{6}(\rho_{{}_M}+2\rho_{{}_R}-2\rho_{{}_{VAC.}})\label{eqn:zz8}
\end{eqnarray}
  Using present cosmological data\cite{Spe03,Fil04,Pea03}, this equation   ultimately implies accelerating universe. Then a reasonable dynamical
age of the universe can be estimated to be $14.2\pm1.7$
Gyr.\cite{Rie98}, consistent with the ages determined by using
various other techniques\cite{Fil04}.
\section{Discussion}
Recent cosmological evidence\cite{Fil04,Ton03} suggests that
cosmological constant  seems to be present evermore in the
cosmological data. Theoretically, cosmological constant appears
when one considers a four dimensional manifold that is due to
compactification\footnote{i.e., using four dimensional
stereographic coordinates.} of a five dimensional manifold having
the signature of a (anti)de Sitter spacetime\cite{Pad03}. This
implies that in the local tangent space the gauge group structure
is either $SO(4,1)$ or $SO(3,2)$. To keep Lorentz invariance
intact  (anti)de Sitter boost is forbidden in the tangent space.
So it is justified, in the present contest, to consider the
Lagrangian as a combination of some $SO(3,1)$ invariant parts of
the full $SO(4,1)$   Pontryagin density.

At first we summaries the main results obtained in this article. These are as follows.
\begin{enumerate}
\item The gravitational constant is related to the NY density by the relation
 $N=-\frac{1}{\kappa}\eta$.
 \item Mass of a localised scalar field $\phi$ is given by the relation $   m_\phi=\frac{c_\phi}{\sqrt{\kappa}}$ where $c^2_\phi$ is the dimensionless $torsion\times\phi$ coupling constant. By this way we get a beautiful analogy of Coulomb's law of electro dynamics in Newtonian gravity. It can be easily checked that, with our previously described  interpretation of mass, the Newtonian force between two gravitating point masses can be written as $\vec{F}=-c_1c_2\frac{\vec{r}}{r^3}$, where $c^2_1$, $c^2_2$ are the two respective torsional coupling constants of the corresponding masses when their dynamics is described by scalar fields in $U_4$.
 \item When $\phi$ represents the local energy parameter at cosmic scale then $\rho_{{}_M}=\frac{\beta}{\kappa}\phi^{2}$, $\rho_{{}_R}=\frac{3\gamma}{2}\phi^{\frac{8}{3}}$ and $\rho_{{}_{VAC.}}=\frac{\lambda}{4\kappa}$. Other kinds of energy densities are disallowed in this scenario. Here, again, $\beta$ is the dimensionless $torsion\times\phi$ coupling constant. Also $\kappa$ together with $\lambda$ are constants of integration.  $\gamma$
is  a  constant having dimension $(length)^{-\frac{4}{3}}$. If $M$ and $V$ be, respectively, the total mass and volume of the universe then $\frac{\beta}{\kappa}=M^2$ and $\rho_{{}_M}=\frac{M}{V}$; this ultimately gives the local cosmological inverse length parameter $\phi=\frac{1}{\sqrt{MV}}$.\end{enumerate}


 It is important to  note  that, in our present formalism, the only
 assumption is that the torsion is represented by an axial
 vector   and the corresponding Lagrangian is a combination of two particular terms
 of the $SO(4,1)$ Pontryagin density in such a way that the $SO(3,1)$
 invariance of the theory is maintained. The presence of the  axial vector at each
space-time point suggests that the space-time manifold is characterized by the presence
of a `direction vector'(vortex line) attached to each point which is the source of torsion.
 It may be remarked that the    degrees of freedom
 of this theory is minimally extended from that of
 Einstein-Hilbert theory with torsion contributing to the
 additional degree. As a result $\kappa$   has got
 its definite geometrical meaning in $U_{4}$ space in
 comparison to  their standard meaning of being simply constants such that  $\kappa$
is   inversely proportional to the  Nieh-Yan density.
      One of the remarkable features of the Lagrangian $\mathcal{L}\mbox{$_{G}$}$ is that
$\frac{1}{\kappa}$  is not a dimensional coupling constant, $\frac{1}{\kappa}$
together with $\lambda$ are constants of integration and they might have
  got there fixed values in the Early Universe just after the bulk
  matter was created when the source of gravity became able to be connected with the
scalar field $\phi$ in the cosmological scale of a FRW-universe. Further the constancy
of $\kappa$ depends upon the form of the source terms in $\mathcal{L}\mbox{$_{G}$}$ such that
these terms are independent of the $SO(3,1)$ connection. Hence separation  of the
tensorial torsion part from the $SO(3,1)$ connection, which is possible only when
the $SO(4,1)$ invariance is broken, and keeping the source independent of the $SO(3,1)$
 connection gives us constancy of $\kappa$.    In other words Lorentz invariance,  in a broken
 de Sitter gauge theory, is associated with the constancy of $\kappa$. This constancy of $\kappa$ also makes it possible to define $mass =\frac{c_\phi}{\sqrt{\kappa}}$ where $c^2_\phi$ is the $torsion\times matter field$ coupling constant. Moreover when we
 consider the metric to have the form of the FRW-cosmology then only three kinds of energy
 densities are possible representing mass, radiation and vacuum energy.  This implies that, in this frame work, other forms of energy densities can be obtained as solutions  when the metric differs from its standard FRW form. It is to be
mentioned here that the scalar field $\phi$ of this paper is different from the Brans-Dicke
scalar field. According to Brans-Dicke
  theory, the value of $G=\frac{c^2 \kappa}{8\pi}$ is determined by the value of the
Brans-Dicke scalar field
  $\phi$. The Brans-Dicke version of Einstein-Cartan theory, with nonzero torsion
  and vanishing non-metricity, was discussed by many  authors\cite{Rau84,Ger85,Kim86}.
   In these approaches $\phi$ acts as a source of torsion\cite{Ber93}. But in our approach
$\phi$ is connected to a local energy parameter. In laboratory scale $\phi$ represents a massive scalar field where the mass arises due to $torsion\times matter$ interaction. In cosmic scale the FRW geometry gives  us $\phi=\frac{1}{\sqrt{MV}}$.

In a recent paper\cite{Mah02} it has been shown that, in the
gravity without metric formalism of gravity, when one performs a
particular canonical transformation of the field variables,
CP-violating $\theta$-term appears in the Lagrangian together with
the cosmological term. This supports the finding of this paper when
we consider that the torsion, being an axial vector, has a certain
role to play in CP-violation. Indeed,   the topological $\theta$-term
 of \textquoteleft gravity without metric formalism' is linked with
 the topological Nieh-Yan density of U$_{4}$
 geometry. In this context we can consider  the finding of
some other work\cite{Ban95}, when the gauge  group is $SL(2,C)$
which is the covering group of $SO(3,1)$, where torsion has been
shown to be linked with CP-violation. Thus arrow of time plays a
significant role in the geometrical origin of torsion and hence of
the gravitational constant. It is to be noted here that the
$\beta$-term, which is the torsion-$\phi$-field interaction in the
Lagrangian $\mathcal{L}\mbox{$_{G}$}$, ultimately gives us the mass-energy
density in (\ref{eqn:22a}) and as $t\rightarrow \infty$ we get
$G_{00}/\kappa\rightarrow$ \textit{the constant energy density of
the de Sitter space} $=\frac{\lambda}{4\kappa}$. Hence our
universe, which is presently accelerating, is heading towards a
universe of constant energy density and infinite radius.

 \noindent
 \section*{ACKNOWLEDGEMENT}
 \vspace{2 mm}

           I wish to thank Prof. Pratul Bandyopadhyay, Indian Statistical Institute,
Kolkata,  for his valuable remarks and fruitful suggestions on this
           problem.

\vspace{4 mm}



             \bibliographystyle{unsrt}
                \bibliography{tbib}

 \end{document}